\def\year{2021}\relax
\documentclass[letterpaper]{article} 
\usepackage{aaai21-arxiv}  
\usepackage{times}  
\usepackage{helvet} 
\usepackage{courier}  
\usepackage[hyphens]{url}  
\usepackage{graphicx} 
\usepackage{natbib}  
\usepackage{caption} 
\usepackage{hieroglf}
\title{Insiders and Outsiders in Research on Machine Learning and Society}
\author{
   Yu Tao\textsuperscript{\rm 1} and Kush R.\ Varshney\textsuperscript{\rm 2}
    \\
}
\affiliations {
    \textsuperscript{\rm 1}Stevens Institute of Technology \\
    \textsuperscript{\rm 2}IBM Research -- T. J. Watson Research Center \\
    ytao@stevens.edu, krvarshn@us.ibm.com
}
\begin{document}

\maketitle

\begin{abstract}
A subset of machine learning research intersects with societal issues, including fairness, accountability and transparency, as well as the use of machine learning for social good. In this work, we analyze the scholars contributing to this research at the intersection of machine learning and society through the lens of the sociology of science. By analyzing the authorship of all machine learning papers posted to arXiv, we show that compared to researchers from overrepresented backgrounds (defined by gender and race/ethnicity), researchers from underrepresented backgrounds are more likely to conduct research at this intersection than other kinds of machine learning research. This state of affairs leads to contention between two perspectives on insiders and outsiders in the scientific enterprise: outsiders being those outside the group being studied, and outsiders being those who have not participated as researchers in an area historically. This contention manifests as an epistemic question on the validity of knowledge derived from lived experience in machine learning research, and predicts boundary work that we see in a real-world example.
\end{abstract}

\section{Introduction}
\label{sec:intro}

Research on the theory and methods of machine learning has led to the ability of technological systems to grow by leaps and bounds in the last decade. With this increasing competence, machine learning is increasingly being employed in real-world sociotechnical contexts of high consequence. People and machines are now truly starting to become partners in various aspects of life, livelihood, and liberty.

This intersection of machine learning with society has fueled a small segment of research effort devoted to it. Two such efforts include research on (1) fairness, accountability and transparency of machine learning (FAccT), and (2) artificial intelligence (AI) for social good. The first of these focuses on the imperative `do no harm' or nonmaleficence, with a special focus on preventing harms to marginalized people and groups caused or exacerbated by the use of machine learning in representation and decision making. The second focuses on using machine learning technologies as an instrument of beneficence to uplift vulnerable people and groups out of poverty, hunger, ill health, and other societal inequities.

\begin{figure}[ht]
\centering
    \includegraphics[width=0.3\textwidth]{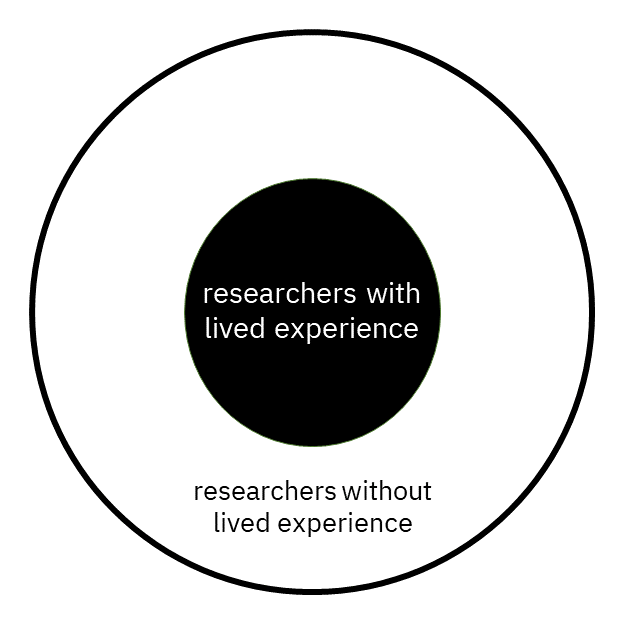}
    \caption{Researchers with lived experience relevant for the topic of inquiry have traditionally been seen as insiders. We hypothesize that the topic of machine learning and society is being conducted at a greater rate by those with lived experience of marginalization.}
    \label{fig:insideroutsider1}
\end{figure}

\begin{figure}
\centering
    \includegraphics[width=0.3\textwidth]{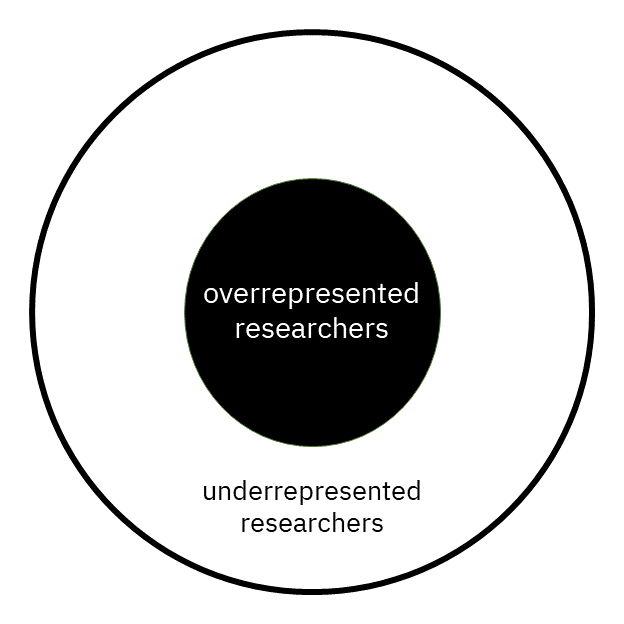}
    \caption{Researchers from overrepresented groups have traditionally been insiders. Machine learning and society is part of a field with underrepresentation of women and racial/ethnic groups.}
    \label{fig:insideroutsider2}
\end{figure}

In this paper, we focus on \emph{who} is conducting this research at the intersection of machine learning and society through the lens of the sociology of science. The theoretical foundation for our investigation is the concept of \emph{insiders} and \emph{outsiders} in the research enterprise \cite{merton1972insiders}. In the social sciences and humanities, researchers are considered insiders if they are members of the community being studied (and thus have lived experience of that community) and outsiders otherwise. (Formal and natural sciences typically do not study communities of people, but the societal aspects of research on machine learning and society does.) A different perspective says that members of groups that have been historically underrepresented in a field of study are outsiders. These two notions, illustrated in Figure \ref{fig:insideroutsider1} and Figure \ref{fig:insideroutsider2} may be at odds. Researchers being insiders from one perspective and at the same time outsiders from the other perspective raises contention in the production of knowledge, including in the epistemic validity of knowledge arising from lived experience. The social construction of whether scientific knowledge arising from lived experience is valid or invalid is an instance of \emph{boundary work} \cite{gieryn1983boundary}.

To analyze researchers in machine learning and society from the theory of insiders and outsiders, first we empirically show that machine learning researchers from underrepresented backgrounds, compared to researchers from overrepresented backgrounds, are more likely to study the societal aspects of machine learning than they are to study aspects of machine learning that are more divorced from society. Recognizing the inadequacy of binary gender categories, we nevertheless take binary gender as one sensitive attribute. (Women are underrepresented and men are overrepresented.) Recognizing the inadequacy of the social constructs of coarse race and ethnicity categories, we also take race/ethnicity as another sensitive attribute. (Blacks and Hispanics are underrepresented, and whites and Asians\footnote{Asians may be disadvantaged in certain considerations like career mobility in a United States context, but are considered overrepresented here in the worldwide machine learning research context.} are overrepresented.) We also examine the intersection of gender with race/ethnicity.

Next, we extrapolate beyond what the empirical analysis is able to tell us by critically examining the factors that may have led to the current state. We also predict the character of the boundary work that may arise in machine learning and society. Finally, through a short case study, we confirm that there is at least  one example in which the theorized epistemic contention has arisen in real life.

The remainder of the paper is organized as follows. After providing a brief recapitulation of research on machine learning and society in Section \ref{sec:mlcscy}, we dive into the theory of insiders and outsiders in knowledge production in Section \ref{sec:status}. In Section \ref{sec:participation}, we discuss the participation of underrepresented groups in science and technology with a focus on computer science. Section \ref{sec:empirical} presents the empirical work; it is conducted on submissions to arXiv, a preprint server that hosts a large fraction of machine learning research papers. Section \ref{sec:discussion} analyzes the sociology of knowledge production in the area of machine learning and society using the theory of insiders and outsiders, and boundary work. We concretize this analysis in Section \ref{sec:casestudy} through a brief case study. Section \ref{sec:conclusion} summarizes and concludes. 

\section{Research on Machine Learning and Society}
\label{sec:mlcscy}

\begin{figure}
\centering
    \includegraphics[width=0.44\textwidth]{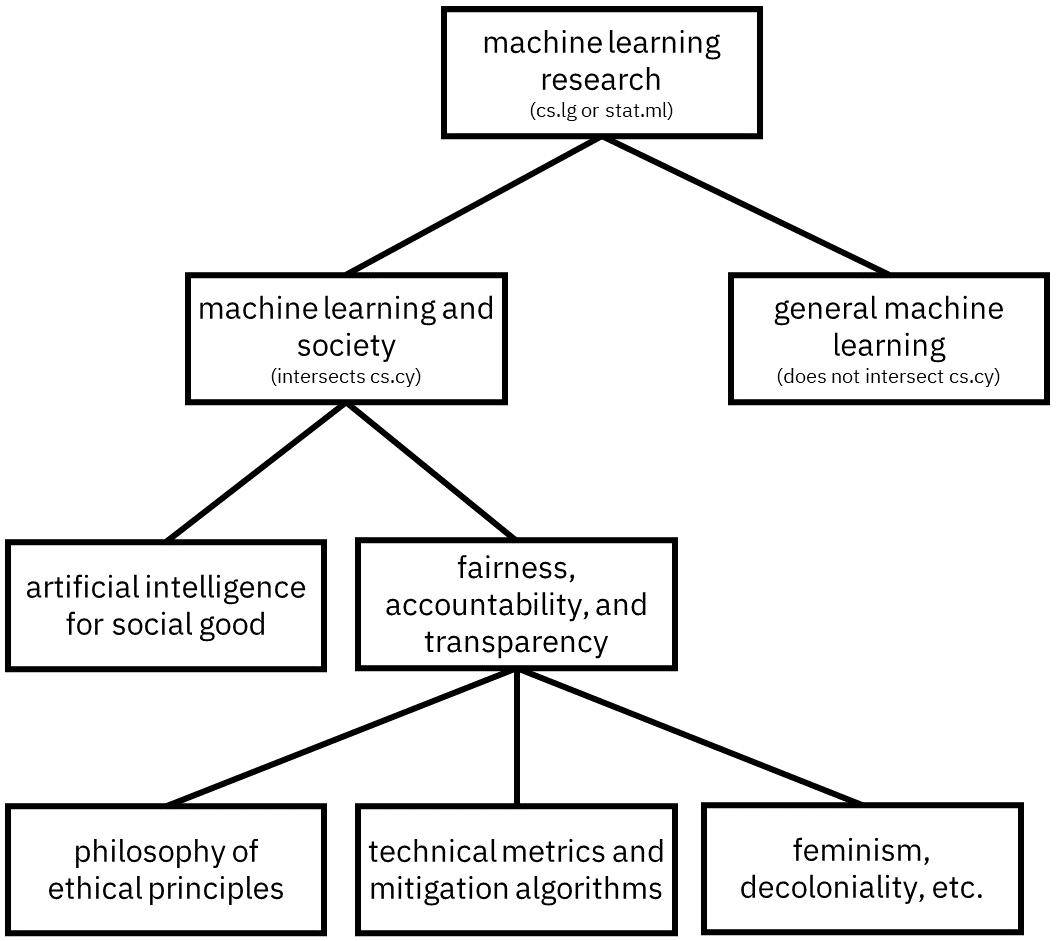}
    \caption{A hierarchical representation of the different topics that constitute research on machine learning and society.}
    \label{fig:topictree}
\end{figure}

As discussed in the introduction, two movements with a societal focus have arisen alongside the growth of research and development of machine learning technologies: FAccT and AI for social good. We briefly summarize these movements in this section, and also in Figure \ref{fig:topictree}.


Ethical AI, responsible AI, trustworthy machine learning, and FAccT all refer to the cross-disciplinary theory and methods for understanding and mitigating the challenges associated with unwanted discrimination, lack of comprehensibility, and lack of governance of machine learning systems used in applications of consequence to people's lives such as employment, finance, and criminal justice \cite{varshney2019trustworthy}. Broadly speaking, there have been three different kinds of research in this area \cite{kind2020the}, including (1) philosophical contributions on ethical principles for AI \cite{jobin2019global,whittlestone2019role}; (2) technical contributions on bias metrics and mitigation \cite{menon2018cost,kearns2019empirical}, explainability and interpretability algorithms \cite{du2019techniques,bhatt2020explainable}, and factsheets as transparent reporting mechanisms \cite{arnold2019factsheets,mitchell2019model}; and (3) contributions bringing forth a social justice angle by adapting theories of feminism, decoloniality, and related traditions \cite{buolamwini2018gender,mohamed2020decolonial}. Research in the first category, ethical principles for AI, tends not to overlap with research on machine learning methods and algorithms. The second and third categories, technical contributions and social justice perspectives, most certainly do intersect with other machine learning research. 

Algorithmic fairness research has two main branches. The first is concerned with allocation decisions like loan approval, pretrial detention judgement, and hiring \cite{barocas2016big}. The program of research is to define mathematical notions of fairness, audit existing systems with respect to those notions, and develop bias mitigation algorithms that optimize for those notions while maintaining fidelity to the learning task. The second branch of algorithmic fairness is concerned with representational issues, for example in information retrieval, natural language understanding, and dialogue systems \cite{blodgett2020language}. Here the program of research mainly revolves around defining the problem itself, since there are many forms of unwanted representational bias ranging from stereotypes encoded into pronouns and occupations, to slurs, offensive language and hate speech, to poorer understanding of dialects and accents of marginalized groups. In both branches, reasons for machine learning models to exhibit systematic disadvantage towards marginalized groups include prejudice of human annotators who label training data, undersampling of marginalized group members in training data, and subjective biases by data scientists in problem specification and data preparation. Research in both branches can span the spectrum from completely formal applied mathematics to wholly social science with calls for justice, i.e., from the second to the third kind of FAccT research. Regardless of where on the spectrum it falls, algorithmic fairness research tends to always be considered part of the machine learning and society nexus.

Explainable and interpretable machine learning, in which the goal is for a person to understand how a machine learning model makes its decisions, has several methodologies appropriate for different contexts and different personas consuming the explanations \cite{hind2019explaining}. One use for explainability is to reveal unwanted biases in machine learning models, but doing so is not reliable \cite{dimanov2020you}. To date, the majority of the research has leaned towards the formal and mathematical. Calls to ground explainability in social psychology and cognitive science \cite{miller2019explanation} have started to bring a greater social science character to the topic. Nevertheless, many interpretability researchers do not consider their methodological work to have a societal aspect, and their papers are not abundant at FAccT-specific venues.  

On the other hand, the framing of efforts to increase transparency of machine learning lifecycles does incorporate a societal angle. For example in factsheets---a tool and methodology for transparently reporting information about a machine learning model as it is specified, created, deployed, and monitored---the reported information can include the intended use of the model as well as quantitative test results on accuracy, fairness, and other performance indicators. It is useful to individuals impacted by the machine learning system (especially those from marginalized groups) and to regulators charged with ensuring the system behaves according to laws and societal values.

Whereas FAccT is concerned with preventing societal harms, AI for social good takes the opposite track and uses the technology to benefit society, especially those at the margins \cite{chui2018applying,varshney2019open}. The working paradigm is to pair data scientists with social change organizations to work towards the 17 Sustainable Development Goals (SDGs) ratified by the member states of the United Nations in 2015, which include: `no poverty,' `zero hunger,' `good health and well-being,' `quality education,' `gender equality,' and twelve others. The form of this pairing may be data science competitions, weekend volunteer events, longer term volunteer-based consulting projects, fellowship programs, corporate philanthropy, specialized non-governmental organizations, innovations units within large development organizations, or data scientists employed directly by social change organizations. Some projects require research and some are more application oriented. The ones that require research and whose results are published fall squarely within the intersection of machine learning and society.

\section{Researchers’ Roles in Knowledge Production: Insider/Outsider Status and Boundary Work}
\label{sec:status}

The insider/outsider discussion in social sciences and humanities addresses the role of the researcher as an insider (i.e., a member of the community being studied) as opposed to an outsider in affecting research, approach, relationship with participants, and/or findings. The insider doctrine of \citet{merton1972insiders} highlights the insider’s exclusive access (the strong version) or privileged access (the weaker version) to knowledge and the outsider’s exclusion from it. Researchers are considered as insiders or outsiders based on their ascribed status (e.g., gender, race, nationality, cultural or religious background) or group membership. The strong version asserts that the insider and the outsider cannot arrive at the same findings even when they examine the same problems; the weaker version argues that insider and outsider researchers would focus on different research questions. The combined version argues that the researcher needs to be an insider in order to understand the community and also to know what is worth understanding or examining about the community \cite{merton1972insiders}. 

However, structurally speaking, it is hard to completely distinguish the insider from the outsider because we all occupy a combination of different statuses, including sex, age, class, race, occupation, and so on. The insider knowledge that is accessible to only individuals who occupy a highly complex set of statuses is limited to a very small group, and this way of knowledge production and sharing is not sustainable. Similarly, social scientists like Karl Marx recognize the value of political, legal, and philosophical theories in economics. Another limitation of the insider doctrine is that it takes a static perspective and does not recognize that our statuses and life experience evolve over time, which shifts our status as an insider or an outsider. In the meantime, the outsider, while not being able to completely transcend existing beliefs and social problems, has the advantages of using less bias in examining social issues and bringing new perspectives to solving issues taken for granted by insiders. The interaction of the insiders and outsiders makes intellectual exchange possible, and Merton argues that we could integrate both sides in the process of seeking truth.

Extending Merton's and other scholars' thoughts on the insider/outsider debate,  \citet{griffith1998insider} also believes that the researcher occupies a particular social location, and her knowledge is situated in particular sets of social relations. However, the insider status is just the beginning but not the end of the research process. Reflecting on her own research experience in mothering work for schooling, she and her collaborator who were both single mothers started as insiders (mothers). However, they had to cross the social and conceptual boundaries to include only mothers from two-parent families (and thus become outsiders in the research process) as the two-parent family is the ideological norm perceived by the schools and society. In other words, researchers are rarely insiders or outsiders but oftentimes insiders and outsiders at the same time, and research is constructed between the researcher and many Others. 

\citet{dwyer2009space} argue that both the insider and the outsider statuses have pros and cons, so what is important is not the insider or the outsider status but ``an ability to be open, authentic, honest, deeply interested in the experience of one’s research participants, and committed to accurately and adequately representing their experience.'' In fact, researchers occupy the `space between.' Challenging the dichotomy and the static nature of insider versus outsider status, the `space between' recognizes the evolving nature of the researcher's life experience and knowledge on the research topic as well as her relationship to participants. 

When there are insiders and outsiders in scientific research, there is also a boundary between them. Specifically, the boundary delineates what is considered `science' and what is considered `non-science' in a particular subfield. Boundary work attempts to shape or disrupt the boundary of what is considered as valid knowledge \cite{gieryn1983boundary,gieryn1999cultural}. Research reveals two types of boundary work: symbolic and social boundaries. Symbolic boundaries are formed when members agree on meaning and definition of the field and obtain a collective identity. Social boundaries enable members' access to material and non-material resources (e.g., status, legitimacy, and visibility) \cite{lamont2002study,grodal2018field}.

For example, \citet{grodal2018field} details how core communities who entered the nanotechnology field early expanded the boundaries of the field by enlarging the definition of the field and associating new members. Peripheral communities, including service providers, entrepreneurs, and university scientists, self-claimed membership during the expansion phase due to newly available material and cultural resources. Later on, while some peripheral communities continued to associate themselves to nanotechnology, the core communities, realizing their collective identity being threatened and resources being restricted because of the enlarged symbolic boundaries, contracted boundaries by restricting the definition and policing membership. Also, some peripheral communities, not identifying strongly with the more restrictive collective identity, self-disassociated and focused on other fields of interest. In this process, the insiders or the core communities  entered the field earlier and had a vested interest to protect, while the outsiders or the peripheral communities entered the field later and had a weak association with the field. The insiders had more power than the outsiders in defining the boundaries of the field and making certain types of work and research legitimate. 

\section{Participation in Computer Science: The Outsider Status}
\label{sec:participation}

\subsection{Participation of Women}

In science, women have been at the “Outer Circle” for a long time. Historically, women faced multiple barriers in entering a scientific career, and even those who were able to become a scientist were not allowed into the inner circles of the emerging scientific community \cite{zuckerman1991outer}. While women’s representation, experience, and advancement in science has increased over time, many of them continue to face barriers, especially at the cultural and structural levels \cite{zuckerman1975women,rosser2004science,hill2010so,national2010gender,ceci2014women}. This is especially true in computer science (CS), where, unlike other scientific fields, women’s participation has been consistently low, with some fluctuations. \citet{hayes2010computer} records the changes of women’s representation in CS in multiple decades: women represented 11\% of all CS bachelor’s degree recipients in 1967; this percentage peaked at 37\% in 1984 and then declined to only 20\% in 2006. For comparison, women represented 44\% of all bachelor’s degree recipients in 1966 and 58\% in 2006, and other STEM fields also witnessed steady increases in this period. Despite the rapid growth of the computer and mathematical science workforce, women’s proportion declined from 31\% in 1993 to 27\% in 2017. However, the silver lining is that among workers with a doctoral degree in these occupations, women’s share increased from 16\% in 1993 to 31\% in 2017 \cite{national2020promising}.

Multiple factors that oftentimes reinforce each other contribute to women's low representation relative to men's in CS at different life stages. Earlier research reports individual factors, such as a lack of early exposure to and experience with computing, women students' inaccurate perceptions of their low quantitative abilities, and a lack of self confidence despite their good performance and computer knowledge level. Other research has also focused on social, cultural, and structural factors which are much harder to change. For instance, women's perceptions of their abilities and the field of computing could be affected by the `chilly classroom' with male students' unfriendly reactions and professors' lack of attention to them; a lack of role models and mentoring; stereotypes against women and against the people, work involved, and values of CS; and the perceived mismatch of women's career orientation to help people and society and what they think CS could offer. Combating these barriers could increase women’s representation in CS or lower their attrition from CS \cite{gurer2002acm,beyer2004deterrents,beyer2006women,cohoon2006just,kim2011engaging,cheryan2015cultural,lehman2016women,cheryan2017some}. 

Policy recommendations and college intervention programs have been made and established to change the cultural and institutional environment in order to recruit and retain more women students and professionals in CS. Some of the recommendations were repeatedly made in different time periods, reflecting a reluctance of change over time. They include involving women students in research at both the undergraduate level and early in their graduate study, actively countering stereotypes and misperceptions of CS, and highlighting and showing women students the positive social impact that scientists can make and the diverse group of scientists making social impacts in their fields \cite{cuny2002recruitment,national2020promising}. Successful college intervention programs in increasing the number and percentage of women CS students and their sense of belonging all tackled the culture of CS and the institution instead of changing the (women) students. These efforts changed the stereotypes of CS by creating introductory CS courses to be inclusive of a diverse student body, providing role models and mentoring to women students, providing research experience, and exposing students to a wide range of applications of CS in solving societal issues \cite{roberts2002encouraging,muller2003underrepresentation,wright2019living,frieze2019computer,national2020promising}. 

\subsection{Participation of Racial and Ethnic Minorities}
In addition to gender, race/ethnicity also shapes scientists’ representation and experience in science as well as their outsider and insider statuses. Among racial/ethnic minorities in a United States context, while Asians tend to be overrepresented in science, the other groups (blacks, Hispanics, and American Indians or Alaska natives) are considered as underrepresented minorities (URMs) due to their low representation in scientific fields, despite their growth over time. For instance, URMs made up 9\% of workers in computer science and mathematics occupations in 2003, and this percentage increased to 13\% in 2017 \cite{khan2020state}. While their participation increased over time, it was still lower than their representation in the general population, confirming their persistent “outsider” status in science. Similar trends hold in a world context with Asians and whites overrepresented compared to black, Hispanic, and indigenous people. 

Research on race and science finds that racial/ethnic minorities, especially URMs, tend to be less likely to publish their research, receive research grants, get recognition for their work, and get promoted but more likely to work in institutions with less resources and more likely to be marginalized in formal and informal scientific communities than their white counterparts. Research also reveals some improvement in their representation in scientific fields as well as in their career experience and outcomes over time, but the progress is slow relative to the growth of the scientific workforce \cite{pearson1985black,pearson2005beyond,ginther2011race,ginther2018using,tao2019gender}. In the meantime, an increasing number of studies employ intersectionality as the research framework that indicates power relations and social inequalities to examine the double disadvantages that minority women scientists suffer from due to both their gender and race, e.g., \cite{malcom1976double,malcom2011double,collins2015intersectionality,metcalf2018broadening}. While minority women scientists of different racial/ethnic groups differ from each other in their career experience and outcomes, they all tend to fare less well than comparable white women as well as men of the same racial/ethnic group \cite{malcom1976double,malcom2011double, pearson1985black,ong2011inside,tao2018earnings,tao2019gender}, revealing the persistent intersectional effect of race and gender. 

\subsection{Status and Career/Research Focus}

Broadly speaking, women\footnote{In this part of the paper and later, we discuss women more than Hispanics and blacks, not because of differing experiences, but because of a dearth of published literature on the analogous experience \cite{spertus1991there}.} tend to be more engaged than their male peers in relatively new, interdisciplinary scientific fields (e.g., environmental studies) that are oftentimes more contextual and problem-based than traditional fields, may not have existing gender hierarchy, and are not well-embedded in the structure of academia or knowledge production, providing more opportunities for women to build the discipline \cite{rhoten2007women}. While some women shy away from technical fields because they do not see the social engagement of these fields, e.g., \cite{carter2006students}, those who choose technical fields do so not only because of the excitement of solving technical problems, but also the potential of addressing issues concerning them and positively impacting people’s lives, which is consistent with their interpersonal and career orientations \cite{silbey2016so, bossart2017women}. Women CS majors choose computing in the context of what they could do for the world with computing---they would like to use the computer in the broader context of education, medicine, music, communication, healthcare, environmental studies, crime prevention, etc. While they also enjoy exploring the computer, the main factor reported by men, women are more likely than their male peers to address the broader social context \cite{fisher1997undergraduate,carter2006students,hoffman2018machine}. While few women in AI were at the “outer circle” in its initial stage, they were attracted to it when it started to develop in the 1980s and 1990s because it was more cognitive than other areas of CS and there were fewer existing stereotypes to fight against \cite{strok1992women}. The intersection of machine learning and society makes careers in machine learning meaningful to them \cite{hoffman2018machine}.

\section{Empirical Study of Knowledge Production}
\label{sec:empirical}

As discussed in Section \ref{sec:participation}, women and URMs tend to select fields in which they perceive they can help people and society. The intersection of machine learning and society provides exactly that opportunity to make social impact. Therefore, we hypothesize that women and URMs are more likely to contribute to research in machine learning and society rather than machine learning without a direct societal component.

\begin{figure}
    \centering
    \includegraphics[width=0.45\textwidth]{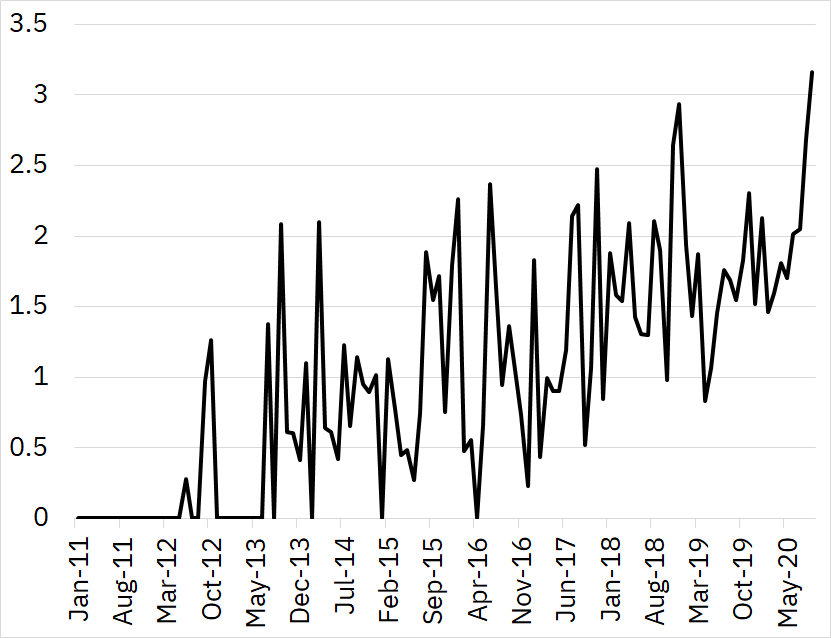}
    \caption{Percentage of machine learning and society papers out of the overall set of machine learning papers posted to arXiv computed on a monthly basis.}
    \label{fig:timeline}
\end{figure}

We performed the following analysis to test our hypothesis. On September 19, 2020, we downloaded the full collection of arXiv paper metadata  available on Kaggle.\footnote{ arxiv-metadata-oai-snapshot.json, available from: https://www.kaggle.com/Cornell-University/arxiv.} The arXiv is a preprint server that has become a de facto standard location for authors to upload their machine learning papers, often irrespective of a paper's publication status in peer-reviewed conferences and journals. When authors upload papers, they self-select one or more subject area categories to tag their paper with. We extracted the subset of machine learning papers among all arXiv papers by filtering those with the set of tagged categories containing stat.ml or cs.lg, the two tags indicating `machine learning.' This gave us 71,605 papers from March 1997 to present. We further marked the papers that also had the tag cs.cy for `computers and society.' We consider these papers to be the ones describing research on machine learning and society. There were 1,077 such papers from August 2004 to present. Thus, the machine learning and society papers represent 1.50\% of the totality of machine learning papers across years. We also analyzed this percentage as function of time, shown in Figure \ref{fig:timeline}. There has been growth over time and the most recent (partial) month, September 2020, had the highest percentage of societally-oriented papers at 3.16\% (except for a few blips in early years when there were small sample sizes). The conclusion thus far is that machine learning and society is a tiny sliver of the overall machine learning universe, but is growing in fraction.

Now, let us move on to analyzing the authors. In contrast to analyses in other sociology of science research, we are not focused on research productivity. Therefore, we did not have to count papers per author or distribute credit among authors. We simply collected the unique set of authors who numbered 103,094. Of these authors, 1,904 had papers only in machine learning and society, 99,460 had no papers in machine learning and society, and 1,730 had papers of both kinds: at least one tagged with cs.cy and at least one other not tagged with cs.cy. 

We used the names of authors to estimate their gender and their race/ethnicity. Such imputation based on observable proxies is fraught with potential biases, as analyzed and discussed at length by \citet{chen2019fairness}. As recommended in that work, we used a soft rather than hard-threshold classifier to minimize the bias as much as possible. For gender classification from the first name, we used the genderize.io API, which has been used frequently for similar analyses \cite{topaz2016gender, hart2019trends,patel2020towards}. If genderize.io was unable to produce a classification, we dropped that author from all subsequent gender analysis. In the analysis that follows, a soft classification score in the range $[0,0.5]$ corresponds to female and a soft classification score in the range $(0.5,1]$ corresponds to male. For race/ethnicity classification, we used two pre-trained neural networks from ethnicolr \cite{sood2018predicting}, which has also been used previously in similar analyses \cite{hofstra2020diversity,chang2020elitism,parasurama2020gender}. We took the simple average of the soft classification score from a model trained on last names from the United States Census and the soft classification score from a model trained on first, middle, and last names from the 2017 Florida voter registrations. Both models have four class labels: Asian, Hispanic, black, and white. While we acknowledge the limitations of such ways of identifying authors' gender and race/ethnicity, they do allow us to analyze the relationship between gender or race/ethnicity and research focus in the absence of better alternatives. 

The average soft classification score of race/ethnicity and of gender, broken down by authorship group membership is shown in Table \ref{tab:all}.
\begin{table}[t]
    \centering
    \footnotesize
    \caption{Average soft classification score of race/ethnicity and gender for three categories of authors.}
   \begin{tabular}{c|c|c|c|c||c}
         & Asian & Hispanic & Black & White & Male \\\hline
        no cs.cy & 0.370 & 0.077 & 0.057 & 0.497 & 0.791 \\\hline
        both & 0.367 & 0.073 & 0.055 & 0.504 & 0.777 \\\hline
        only cs.cy & 0.266 & 0.097 & 0.071 & 0.566 & 0.726 \\\hline\hline
        slope & -0.0430 & 0.0077 & 0.0055 & 0.0298 & -0.0293 \\\hline
        p-value & 0 & 0.0062 & 0.0241 & $<$0.0001 & $<$0.0001 \\\hline
    \end{tabular}
    \label{tab:all}
\end{table}
Recall that the three rows in the table, the author groups, correspond to authors having no machine learning papers with a societal aspect, authors having machine learning papers both with and without a societal aspect, and authors having machine learning papers only with a societal aspect. First look at the gender column labeled `Male.' (It is labeled as such because the number indicates a (possibly uncalibrated) probability of being male.) The numbers decrease as we go down the table, which indicates that the overrepresented male group is less likely to work on machine learning and society and the underrepresented female group is more likely to work on machine learning and society. The table also gives the p-value and slope for the trend in proportion obtained using the Cochran-Armitage trend test. The negative slope result is consistent with our hypothesis. 

Next focus on Asians; again, the slope is negative, and again, an overrepresented group is less likely to work on machine learning with societal impact. In the next two columns, Hispanic and black, the trend is reversed. This indicates that these two underrepresented groups are more likely to work on machine learning with societal impact. These results are statistically significant and consistent with our hypothesis. 

Finally, if we look at the white column, we see the numbers increase (positive slope), which is the hypothesized trend of underrepresented rather than overrepresented groups. To understand this better, we conducted an intersectional analysis of race/ethnicity by hard-thresholded gender, presented in Table \ref{tab:male} and Table \ref{tab:female}.
\begin{table}[t]
    \centering
    \footnotesize
    \caption{Average soft classification score of race/ethnicity among estimated males for three categories of authors.}
    \begin{tabular}{c|c|c|c|c}
         & Asian & Hispanic & Black & White  \\\hline
        no cs.cy & 0.335 & 0.078 & 0.059 & 0.528  \\\hline
        both & 0.343 & 0.076 & 0.057 & 0.525  \\\hline
        only cs.cy & 0.247 & 0.097 & 0.073 & 0.583\\\hline\hline
        slope & -0.0345 & 0.0072 & 0.0051 & 0.0223 \\\hline
        p-value & $<$0.0001 & 0.0113 & 0.0404 & $<$0.0001 \\\hline
    \end{tabular}
    \label{tab:male}
\end{table}
\begin{table}[t]
    \centering
    \footnotesize
    \caption{Average soft classification score of race/ethnicity among estimated females for three categories of authors.}
    \begin{tabular}{c|c|c|c|c}
         & Asian & Hispanic & Black & White  \\\hline
        no cs.cy & 0.446 & 0.071 & 0.050 & 0.433  \\\hline
        both & 0.420 & 0.067 & 0.051 & 0.463  \\\hline
        only cs.cy & 0.287 & 0.093 & 0.068 & 0.551\\\hline\hline
        slope & -0.0695 & 0.0081 & 0.0077 & 0.0537 \\\hline
        p-value & 0 & 0.0029 & 0.0007 & 0 \\\hline
    \end{tabular}
    \label{tab:female}
\end{table}
Looking at the white column in both tables, we see that the increase is much stronger for white females than white males, which again is consistent with the hypothesis of underrepresented groups having an affinity for making social impact. The Hispanic and black columns in these two tables are similar to the aggregate result in Table \ref{tab:all}. The Asian columns for males and females have differences, however. Asian females show a stronger trend than Asian males in the direction hypothesized for overrepresented groups. 

To reconfirm this finding, we also did the intersectional analysis in the opposite direction: by taking the hard classification of the predicted race/ethnicity and producing the Tables \ref{tab:asian}--\ref{tab:white}. 
\begin{table}[t]
    \centering
    \footnotesize
    \caption{Average soft classification score of gender among estimated Asians for three categories of authors.}
    \begin{tabular}{c|c}
         & Male   \\\hline
        no cs.cy & 0.738   \\\hline
        both & 0.732   \\\hline
        only cs.cy & 0.696 \\\hline\hline
        slope & -0.0183 \\\hline
        p-value & $<$0.0001 \\\hline
    \end{tabular}
    \label{tab:asian}
\end{table}
\begin{table}[t]
    \centering
    \footnotesize
    \caption{Average soft classification score of gender among estimated Hispanics for three categories of authors.}
    \begin{tabular}{c|c}
         & Male   \\\hline
        no cs.cy & 0.830   \\\hline
        both & 0.815   \\\hline
        only cs.cy & 0.722 \\\hline\hline
        slope & -0.0473 \\\hline
        p-value & 0 \\\hline
    \end{tabular}
    \label{tab:hisp}
\end{table}
\begin{table}[t]
    \centering
    \footnotesize
    \caption{Average soft classification score of gender among estimated blacks for three categories of authors.}
    \begin{tabular}{c|c}
         & Male   \\\hline
        no cs.cy & 0.825   \\\hline
        both & 0.871   \\\hline
        only cs.cy & 0.721 \\\hline\hline
        slope & -0.0345 \\\hline
        p-value & 0 \\\hline
    \end{tabular}
    \label{tab:black}
\end{table}
\begin{table}[t]
    \centering
    \footnotesize
    \caption{Average soft classification score of gender among estimated whites for three categories of authors.}
    \begin{tabular}{c|c}
         & Male   \\\hline
        no cs.cy & 0.822   \\\hline
        both & 0.801   \\\hline
        only cs.cy & 0.739 \\\hline\hline
        slope & -0.0375 \\\hline
        p-value & 0 \\\hline
        
    \end{tabular}
    \label{tab:white}
\end{table}
The average soft classification score for gender (probability of male) is much lower for Asians in the no societal impact category, shown in Table \ref{tab:asian}, than all other race/ethnicity groups, shown in Tables \ref{tab:hisp}--\ref{tab:white}. Asian females work on machine learning research with a societal angle the least among the different female racial/ethnic groups. Broadly speaking, the results are consistent with the hypothesis that underrepresented groups defined by Hispanic or black race/ethnicity or female gender focus more research efforts towards machine learning and society than machine learning without societal impact.

\subsection{Limitations}
The empirical methodology has limitations, but still provides us with a unique opportunity to understand the relationship between insider/outsider status and research focus. As mentioned before, using proxies for sensitive attributes has limitations in general \cite{chen2019fairness}, but more so when the proxy is the person's name. The name does not accurately identify the gender and/or race/ethnic background equally well for all people, with a particular confusion between Hispanics and white southern Europeans. Also, the use of arXiv as a de facto standard repository for machine learning research was not true in early years. We minimize this concern by only \emph{comparing} machine learning papers with and without a societal aspect, rather than looking at absolute numbers. There may be some small bias in the analysis if authors of works with and without societal aspects behave differently in arXiv posting behavior, which may happen because of different publication practices in the social sciences and humanities compared to CS.

\section{Theoretical Analysis}
\label{sec:discussion}

The insider/outsider discussion is embedded in qualitative research in social sciences and humanities, and it has addressed the impacts of the researcher’s insider or outsider status on the role they play in their research, participants’ trust of and openness to the researcher, and the data being collected, e.g., \cite{labaree2002risk,mercer2007challenges,kerstetter2012insider,hayfield2015insider}. However, this discussion has an implication on research on machine learning and society as well. On the one hand, while women and URM machine learning researchers do not necessarily have face-to-face interactions with their subjects like some social scientists, they do have lived experience of unwanted bias in society and the workplace. They care about redressing social inequalities and can bring that insider experience into their research. This insider status is illustrated in Figure \ref{fig:insideroutsider1}. On the other hand, women, Hispanics and blacks are historically underrepresented in science and engineering fields, especially CS, which makes them outsiders, illustrated in Figure \ref{fig:insideroutsider2}. It is worth noting that this outsider status does not refer to a lack of technical competence, it only means that they are underrepresented in the ML community and, as a result, it is hard to get into the inner circle of the community. However through the empirical study of Section \ref{sec:empirical}, we know that women and URMs are overrepresented in research on machine learning and society as compared to plain machine learning research.

\subsection{Analysis of the Current State}
Despite being at the center of building the field of machine learning and society research, women’s (and URM's) experience in the workplace reflects their overall struggles in society. Similar to women in some other scientific fields, women computer scientists tend to be more subject to stereotyping, less likely to be full professors or in senior research and technical positions, less recognized for their work and paid less, more likely to be subject to overt discrimination and harassment, more likely to face pressures in balancing work and life, and more likely to be marginalized than their male peers \cite{strok1992women,simard2010senior,rosser2004science,tao2016they,fox2016women,khan2020state}. These ``outsider'' disadvantages provide them with the insider perspective when conducting algorithmic fairness and other socially-oriented machine learning research. In fact, women (and URM) scientists' lived experience and consequent insider status place them in a unique position to formulate questions and conduct research at the intersection of machine learning and society. 

The finding that women and underrepresented minorities are more likely to work on machine learning and society research should not be interpreted as that all insiders conduct only machine learning research without the social aspect and all outsiders conduct only machine learning and society research. However, this finding is consistent with literature that reveals women' and underrepresented minorities' preference for conducting and applying research in a broader context---one that goes beyond the technical. As insiders of social inequality, they bring their lived experience and the new perspective into a field where they have been outsiders. Now in the late 2010s and early 2020s, machine learning and society represents an area of AI that is relatively new, interdisciplinary, not well-embedded in the structure of academia, and without existing hierarchies, and thus one with an opportunity for women and URMs to build, which they are doing.

While our empirical analysis finds that women of different racial/ethnic groups tend to behave more similarly to each other than to their male counterparts, we also find that the women's groups differ from each other, confirming the intersectionality perspective and that some groups are not purely insiders or purely outsiders. We would like to highlight Asian women, who are in a unique position in machine learning (or science as a whole) because Asians are overrepresented but women are underrepresented in science. Asian men tend to behave similarly to their white counterparts in their career outcomes, but Asian women tend to behave more like other women's groups, making the gender gap among Asians greater than that among some other racial/ethnic groups \cite{tao2015engineering,tao2018earnings,tao2019gender}. Being insiders in machine learning on the one hand (Asians) and being outsiders on the other hand (women) could possibly constrain some of their choices because they may receive inconsistent expectations and experience multi-level barriers. In the meantime, the Asian and Asian American cultures tend to emphasize technical expertise and the instrumental value of education to fight their marginal status and to achieve upward social mobility in American society \cite{xie2003social,min2015concentration}. The emphasis on technical aspects and some structural barriers they experience in their careers may suggest that Asian women pursue machine learning occupations due to the technical and financial aspects more than the social impact of such occupations that are more likely to be highlighted by other women's groups.

In addition, the findings reveal complicated issues of power and inequality in the ML community, which reflects societal inequality. Both at the personal level, e.g., in terms of exposure to and experience with computing at an early age, and at the cultural and structural levels, e.g., in terms of experiences in computing classes and workplace, statuses (e.g., as women or racial/ethnic minorities) affect our lived experience and opportunities to pursue a career in science. When entering science, our lived experience could impact our research focus. While women and URMs are not outsiders to ML in the sense of being less technically competent, they are outsiders as historically underrepresented groups that have not been successful in penetrating into the inner circle. As a result, they are not in a position of power but are disadvantaged in various ways. In the meantime, being insiders to the experience of inequality, they use their technical expertise to address and provide solutions to persistent social inequality. In this sense, they are empowering not only themselves as underrepresented groups but also the ML community by raising awareness and impact of ML research with social implications. 

\subsection{Epistemic Conflict}
Outsiders' entrance into the field could be shaped by existing barriers and policed by the insiders. Once outsiders enter a field, they have another challenge of making legitimate the research that they prefer but somehow diverge from the mainstream. Although research driven by lived experience (including the third category of FAccT research that brings in feminist, post-colonial, and other related critical-theoretic thoughts) may be celebrated within the intersection of machine learning and society, it is questioned outside of the intersection on epistemic grounds. According to \citet{haraway1988situated}, knowledge is situated and embodied in specific locations and bodies, and the multidimensional and multifaceted views and voices, from both those in power and those with limited voices, combine to make science. Nevertheless, despite scholarship supporting lived experience not being in conflict with scientific objectivity, the common refrain summarized by the feminist and postcolonial epistemologist Sandra Harding is as follows: ```Real sciences' are supposed to be transparent to the world they represent, to be value neutral. They are supposed to add no political, social, or cultural features to the representations of the world they produce.'' In other words, ways of knowing that do not follow the (Western) scientific method are not seen by practitioners as scientific \cite{harding2006science}, despite scholarly criticisms to this perspective. The implication in the context of machine learning and society is that critical-theoretic work based on lived experience as the source of knowledge will be discounted in mainstream machine learning: insider research by outsiders is precarious.

\subsection{Boundary Work Predictions}

Based on the sociology theory, we may predict two possible futures, both involving \emph{boundary work}. The first is a severing of the connection between mainstream machine learning research and societally-relevant applications and governance, i.e., the \emph{expulsion} of machine learning and society from mainstream machine learning. The second is the \emph{expansion} of machine learning research to include knowledge from lived experience, while overcoming tendencies for expulsion and the protection of autonomy that many insiders of machine learning research may have.

The future that emerges among the two possibilities could depend on what insiders perceive as legitimate, as in the case of nanotechnology. (The insiders hold epistemic authority both due to their entrenched status and the power that comes from their identity (race/ethnicity, gender, etc.) \cite{pereira2012feminist,pereira2019boundary}.) In addition, another factor may shape the future of machine learning and society research: sustainability of the ML field as a thriving site of research to continuously attract the next generation of scholars, including women and underrepresented minorities. When the outsiders conduct their machine learning and society research, they raise awareness of the broader context of technical issues. While machine learning and society research is still a small portion of machine learning research overall, it has been growing. Led by ``the outsiders,'' this line of inquiry is increasingly being addressed and published. Based on this trend and considering that the field could benefit from both insiders and outsiders' perspectives, we have reasons to believe that machine learning and society research will transform and sustain ML knowledge and practice, even though it may not happen soon and there may be backlashes. 

\section{Case Study}
\label{sec:casestudy}

Let us see if our boundary work predictions from Section \ref{sec:discussion} hold in a specific case study. In June 2020, the software for an image super-resolution algorithm \cite{menon2020pulse} was posted on GitHub and soon discovered to alter the perceived race/ethnicity of individuals whose downsampled face images were presented as input \cite{johnson2020ai,kurenkov2020lessons,vincent2020what}. Examples of input black, Hispanic, and Asian face images yielded white-looking results.  About these results, Facebook machine learning researcher Yann LeCun commented on Twitter: ``ML systems are biased when data is biased. This face upsampling system makes everyone look white because the network was pretrained on FlickFaceHQ, which mainly contains white people pics. Train the *exact* same system on a dataset from Senegal, and everyone will look African.'' 

In response, Google machine learning researcher Timnit Gebru pointed to the video of her recently completed tutorial (with Emily Denton) \emph{Fairness Accountability Transparency and Ethics in Computer Vision} with the comment: ``Yann, I suggest you watch me and Emily’s tutorial or a number of scholars who are experts in this are. You can’t just reduce harms to dataset bias. For once listen to us people from marginalized communities and what we tell you. If not now during worldwide protests not sure when.'' She also posted: ``I’m sick of this framing. Tired of it. Many people have tried to explain, many scholars. Listen to us. You can’t just reduce harms caused by ML to dataset bias.'' A back and forth debate ensued on Twitter with many interlocutors taking sides and offering inputs. 

Let us analyze what happened using the insider/outsider understanding of research on machine learning and society that we have developed in this paper. LeCun is a white male, Chief AI Scientist at Facebook, and Turing award winner---a person likely without lived experience of marginalization and a clear insider in mainstream machine learning research. Gebru is a black female, co-lead of the Ethical Artificial Intelligence Team at Google at the time---a person with lived experience of marginalization and thus an insider in algorithmic fairness research, but an outsider in machine learning overall. 

Although \citet{gebru2019datasheets} say: ``Of particular concern are recent examples showing that machine learning models can reproduce or amplify unwanted societal biases reflected in datasets,'' which is consistent with LeCun's argument, Gebru's comments in the debate point to her holding a stance consistent with Merton's insider doctrine of lived experience providing privilege (bordering on exclusivity) for conducting research on machine learning bias. Additionally, her epistemic perspective appears to be that such lived experience is a valid source of knowledge for ``many scholars.'' The repeated call to listen to scholars is an attempt at expansion boundary work. On the other hand, LeCun's perspective epitomizes a boundary and epistemic authority that leaves lived experience out of machine learning research; scientifically-derived knowledge is valid knowledge. At the end, some white males in positions of power also joined the debate and offered allyship, which may have expanded the epistemic boundary of machine learning just a little bit. What may have appeared at first glance to be a personal war of words was in fact an example of boundary work in practice, manifested as contention between two insiders of their own respective domains.

After we completed the first draft of this paper in October 2020, there was further contention involving Timnit Gebru in December 2020 \cite{johnson2020google}. She was dismissed from her research position at Google by Jeff Dean and Megan Kacholia, who are both white and in positions of power. The dismissal was widely argued in a public manner. Among others, one of the factors was Gebru's reluctance to remove Google authors from or withdraw the paper ``On the Dangers of Stochastic Parrots: Can Language Models Be Too Big? $^{\tiny{\textpmhg{\Ha}}}$'' \cite{bender2021danger} from the ACM FAccT Conference. One of the main parts of this paper conducts a critical analysis of large language models through the lens of decolonizing hegemonic views \cite{srigley2018decolonizing}, which is a prototypical example of the third, social justice, angle to FAccT research  mentioned in Section \ref{sec:mlcscy}. Dean and Kacholia's criticism of the paper was the inclusion of the decolonial perspective at the expense of a technical-only analysis that would include a discussion of techniques to mitigate representation bias, which correspond to the second kind of FAccT research mentioned in Section \ref{sec:mlcscy}. The first author of the paper, Emily Bender, posted on Twitter: ``The claim that this kind of scholarship is `political' and `non-scientific' is precisely the kind of gate-keeping move set up to maintain `science' as the domain of people of privilege only.'' This case is also one of epistemology and further illustrates how boundary work can engender extreme conflict and the tendency for expulsion boundary work.

Together, these cases illustrate a little bit of expansion boundary work and a healthy dose of expulsion boundary work, both as predicted by the theory of insiders and outsiders.

\section{Conclusion}
\label{sec:conclusion}

In this paper, we have studied the sociology of researchers creating new knowledge in the area of machine learning. We have analyzed the intersection of general machine learning research with FAccT and AI for social good---collectively machine learning and society---using Merton's concepts of insiders and outsiders. Although these concepts are usually only applied in studying the social sciences and humanities rather than the natural sciences and formal sciences, there is a clear insider in terms of lived experience in machine learning and society: a member of a marginalized group. Just as importantly, these same marginalized groups have low historical participation and representation in computer science, the parent field of machine learning and are thus outsiders in a different way. Through an empirical study, we find that researchers from marginalized groups are overrepresented in conducting research on machine learning and society. Therefore, we have a situation in which the same group takes the insider role in terms of lived experience and the outsider role in terms of historical participation. This situation leads to tension over the boundaries of valid knowledge in machine learning and society. Specifically, the epistemic question that arises is whether lived experience is a valid source of knowledge. Instances of expansion and expulsion boundary work are predicted and verified in a case study.

If one takes the normative stance that expansion boundary work is preferable to expulsion, then the resolution of the epistemic contention calls for facilitation by researchers with ascribed status that is not marginalized and who lean towards including knowledge derived from lived experience within the boundary of machine learning research.

This paper illuminates a few avenues for future research. One such direction is to dive deeper into the topics of situated knowledge, feminist epistemology, and boundary work to better understand how the field of machine learning and society may evolve and to understand strategies for directing that evolution in a beneficial way. Another future direction is to study the impact of papers in machine learning and society produced by teams containing only members with lived experience with marginalization, containing only members without lived experience with marginalization, and containing both types of researchers, to understand whether work that bridges the epistemic divide of formal science and critical theory is more valuable than other pieces of work.

\section{ Acknowledgments}
The authors thank Delia R.\ Setola for her assistance and Tina M. Kim for her comments.

\bibliography{outsiders}

\end{document}